# Bio-inspired Political Systems: Opening a Field

Nathalie Mezza-Garcia

Political Science Department,
Universidad del Rosario,
Bogotá, Colombia
`meza.nathalie@ur.edu.co`

*Extended Abstract*

In this paper I consider the necessity and the possibilities of engineering *bio-inspired political systems*. Systems capable of harnessing the complexity of the human social systems and their increasingly growing and diversifying interactions.

Political science has merged complexity, until a certain degree, but there still is a long way to go. Although few authors have properly used tools, phenomena and concepts regarding complexity such as non-equilibrium thermodynamic [1], chaos theory [2], evolution [3], game theory [4], self-organization [5], entropy [6], modeling and simulation [7], among others, so far the majority of studies are still discussing in terms of elections and voting dynamics [8], political parties [5], the nation-state [9], representative democracy [10], public policies [11] and other models of organization typical from our current political structures, which, as it happens with actual political systems, are also taken for granted. This is because no one has taken radically and seriously the living and increasingly complexifying nature political systems are starting to gain, as human social systems maximize their degrees of freedom and expand their possibilities, using Stuart Kauffman´s expression, through the adjacent possible. Political systems haven´t even been part of the study of the philosophy of complex systems [12]. In most of the studies it is considered that the past was a succession of necessary events that led us to the best, actual and last way to politically organize our human social systems. As if there was no future. Here I want to go beyond that stationary conception. For that reason I introduce a tendency whose concept hasn´t been developed until now: *bio-inspired political systems*.

By political systems, I will refer to political power configurations. *Id est,* political systems as political science traditionally understands them. Although mine is a conceptual study, I will assume political science´s concepts as the current mainstream of science understands them. The study is based upon a historical perspective, however, from this approach, I don´t consider tribal forms of political systems by virtue of their variety, the rich differences between them and because none of them was extensively hegemonic in space. Nonetheless, since my concern is the future, from this standpoint, tribal organizations will be included in the study of the new relations, structures and modes of organization of the political systems, in behalf of that same diversity and their less artificial and imposed organizing structures. As it´s known, political systems may vary and differentiate one from another in their regimes, such as parliamentary or presidential ones; in their expression in political modes of territorial organizations, like ancient, feudal or modern states; and in having very different institutions. Additionally, their models can –apparently- be far from one and other depending on the space and time where they were implemented. Yet, the term *political system* encompasses any political model with an implicit notion of managing power.

Having said that, I conjecture that engineering *bio-inspired political systems* can be useful when thinking political systems as we approach less imposed and more natural modes of organizing our human social systems. Bio-inspired political systems will manifest, at first, in the form of self-organized control networks, just as it is in our biological referent [13] and they will continue to evolve towards networks without the need of control. I base on a comprehension of life and its organic properties to define them as political systems where synthesis processes in individual´s complex networks interactions self-organize human social systems with swarm intelligence. Bio-inspired political systems will be a consequence of the non-linearization of human relations and their interactions. They entail a previous complexification of human social systems. Their emergence will occur when top-down centralized political systems stop being able of framing or containing the social phenomena, tendencies or individual´s attitudes that come with exponentially augmented information flows. This will happen when personal or collective identity can no longer be traced back by the adscription to a determined group or community and when tags can´t easily be placed just by considering adscription factors. Even less when relations between individuals' identity marks are covered or understood separately.

*Bio-inspired political systems* will be the mechanisms permitting the development of non-imposed economic and social structures, instead of blocking them from happening as our top-down artificial political systems have done so far. They are the best way, I claim, to properly stepping on the ground of the future emergent and complex social processes that lie ahead. They are also the only mechanisms a true direct and real democracy can exist in practice, if the possibilities to participate and decide in the «public arena» continue to be a central issue in humans' relations in the future. This, in contrast to our actual rudimentary representative democracy.

The certainty about this tendency derived from the *longue durée* approach [14] of analysis I used as a starting point for the research. It reveals when diacronically considered that the exponential growth of proliferation of interactions will maintain its tendency as long as we continue to increase the amount of energy we consume, no matter what kind of energy it is we are extracting. Historical and technological evidence shows that the proclivity, despite fluctuations, goes in that direction; thereafter, one can safely say that the complexity of human social systems will not cease to expand and diversify through out the adjacent possible [15]. This implies that the spectrum of individual identities will amplify the options related to the link between personal identity and features such as territorial adscription, migration, labor specialization, nationality, sexual choices, political inclination, religion, everyday likes, trends and so on. This dissociation definitely has an impact on political systems and connects with my thesis of bio-inspired political systems as the political systems of the future. It also raises the concept of *political granularity*, a term I extrapolated from molecular dynamics [16] and engineering [17].

The background of my thesis is a *critique* to social relations in centralized hierarchical political structures, which is a shared characteristic of all the hegemonic political systems that have been artificially implemented since human recorded written history in non-tribal societies. By hegemony we mean those political systems whose power relations, structures and modes of organization where dominant in space for a period of time exceeding the basic number of units of time in political science –namely the year- than the average life expectancy of humans at the moment in space and in time where they were artificially implemented. This does not implies that during their phase transitions they were mutually exclusive [18]. As we´ll see when considering the political granularity.

For my purposes I define *political granularity* as a category for defining the extension of the territorial parcels on which modes of organization are imposed depending on the specific political system that is superimposed over the human social system that remains in that territory. I refer to these parcels as grains, and their extension defines their granularity. The States' territories, provinces, regions and cities are examples of grains. Since the beginning of written history there is a tendency of progressively non-linear finer administrative and territorial political granularity, going from coarse-grained parcels to fine-grained parcels. This is so, even though political coarse grains are formed, at first, from political finer grained grains; i.e. the EU is conformed by states or, from another point of view, big empires were conformed by smaller territorial organizations [19]. I support my thesis in the idea that although it is highly possible that coarse-grained political grains emerge as modes of organization from particular political systems, reuniting fine-grained political grains, they will no longer be able to mobilize as much power as they have had up to-date. One of the reasons that explains the decrease in their capacity to concentrate power is that simultaneously with the complexification of human social systems as a consequence of the complexification of human social networks and interactions, we are currently losing the notion of spatiality. Local exchanges are passing from having defined territorial or geographic spaces to not having a notion of territorial space at all.

Figure 1. is an example of political granularities over the past 2500 years approx. As grains we consider modes of organization defined by three different political systems: thus, oligarchy (empires), monarchy (kingdoms) and democracy (states). We can see how fine-grained political parcels –states- emerged from coarse-grained political parcels –kingdoms- and how these last ones arose from even more coarse-grained parcels.

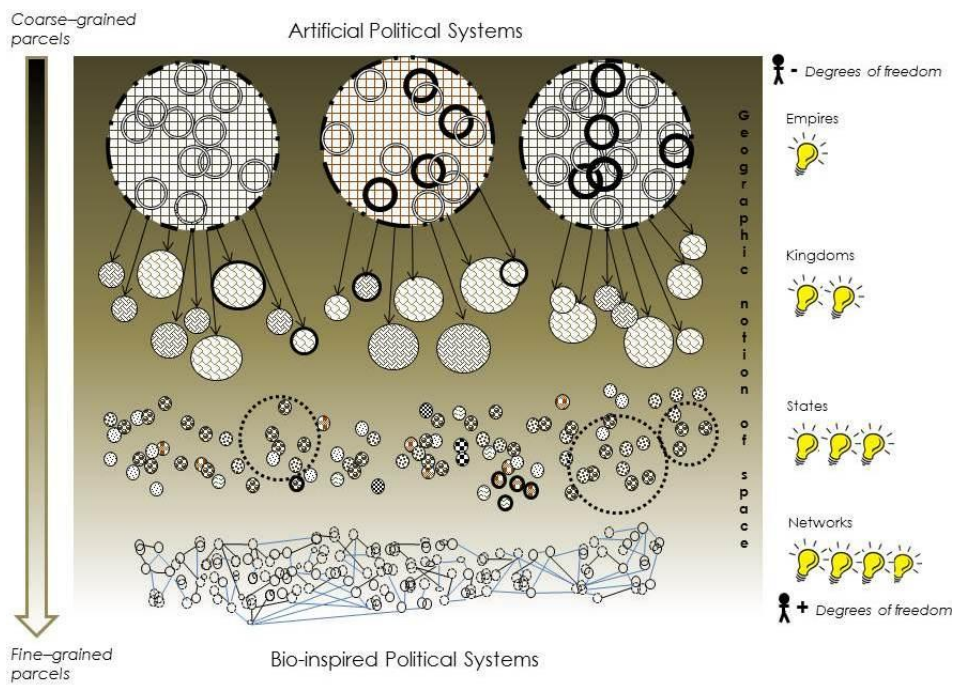

**Figure 1. Political granularity. (Source: the author)**

We can also see how there is something similar to re-unifications presented in the phenomenon of supranational organizations, as the EU could be, for instance. However, it is important to acknowledge and compare the circumstances where this reunited supra-entities emerged with the actual social and informational context in which territorial or geographic space lose importance.

There is, in fact, a positive feedback loop between the every time finer political granularities, unlinking personal identities with a specific territory and the loss of the notion of territorial space as the *arena* were interactions take place. Consequently, *bio-inspired political systems* are rather modes in which societies control their self-organization than actual power distribution systems. Indeed, anarchism, understood as a political system, would be the only one left out from that consideration since it does not consider the imposition of any form of organization over a geographical area. Anarchism is entirely compatible with the idea of the self-organization of networks due to its focus in non-centralized control [20] and it´s in harmony with the most probable, possible and plausible economic model that will accompany the phenomenon here introduced: catalaxies. Catalaxies refers to a spontaneous order in economy resulting from individual specialized exchanges [21]. Inevitably this understanding leads towards more organic political systems. We are experiencing an inflection point from the old mechanistic tradition.

A critical point is that, as Prigogine and Stengers said [22], we are experiencing the end of scientific and social deterministic models when dealing with social phenomena. Our social systems are likely to have increasingly emergent properties due to interactions in self-organized processes in bottom-up complex networks and our political systems must reflect that tendency and learn to harness it. The structure of the nation-state is a simplifying and reductionist approach that can´t properly manage the tendency of the complexification of complex human networks because of its top-down anti-emergent properties. The centralized top-down expected cause-effect relations in political system´s tree network topology says nothing about the social relations going under, behind and beyond those political structures, in contrast to the non-linear dynamics and the decentralized scale free network topology of human interactions in social systems. Punctually, our actual static political systems know little, or nothing, about the increasingly complex nature of the human social systems on which they are imposed [23]. We already have the technological means to engineering bio-inspired political systems plus individuals' will to participate, as internet´s social networks have shown. We are just missing less intrusive, yet poorly effective, political systems.

It might sound as a contradiction to refer to engineering political systems and let them self-organize. However, we don´t mean e*ngineering* in the traditional reductionist and deterministic sense. We are talking about complex engineered systems. Bio-inspired engineering is one way to show the complexification of engineering [24] and the complexification of our world, therefore bio-inspired political system´s engineering must emphasize methods that involve that complexification. In this context, the best way is to combine metaheuristics and simulation (notably agent-based modeling and simulation). Together, they can contribute to a better comprehension, understanding and treatment of the phenomena. Metaheuristics lead to finding better solutions, whereas simulation allows to make concept proves, something that political science can´t do in the real world. In agent-based simulations I have implemented, I managed to synthetize self-organized control mechanisms, which suggest that engineering bio-inspired political systems without the need of control is plausible. This results, however, aren´t the focus of this exploratory introductory paper. They are the subject of a work in progress.

One possible argument against the need in engineering bio-inspired political systems could be that we have, in some senses, progressed despite the imposition of our deterministic, centralized and static political systems and their structures. However, the complexity of human social systems long time ago can´t be compared with their complexity nowadays [25], therefore we can´t continue with the same kind of relations in our political systems' structures. Apart from that, our artificial political systems with their entangled economic structures have had harmful consequences for many humans and other forms of life in the planet. That said, the ethic scopes of engineering bio-inspired political systems are conclusive. Less anthropocentric and artificial ways of organizing and structuring our human social systems will influence the way we exploit Gaia´s [26] resources, evolving towards more dynamic equilibriums between our human social systems, natural and artificial social systems; undoubtedly the greatest challenge ever faced in human history [27]. From this standpoint, *bio-inspired political systems* signify self-organized networks of, likewise, self-organized networks conformed by individuals, processes, learning dynamics, resources, information, etc. adapting to achieve dynamic equilibriums among the human social systems and in their relation with natural and artificial ones. They imply parceling the globe´s territory and management in finer-grained political grains, reducing the capacities of *mainframe* power concentrators. An essential point is to choose which are the elements in the base of human social systems that give rise to adaptive intelligent human social systems and modes of interactions in a world loosing it´s focus in territory.

Biological systems become our best reference to understand the behaviors that underline the phenomenon and the phenomenon itself. Modeling and simulation added to metaheuristics, plus a historical analysis gave us and will give us a better understanding of the adjacent possible of actual political systems. It can be said that bio-inspired political systems signify that Plato´s politeia becomes a reality, a living-constantly-changing entity. I´ll expand this idea farther in the paper. Again, I insist: *bio-inspired political systems* are the only way in which, if political systems continue to exist, a political system can harness the complexity of human social systems and their interactions. It´s time for political science to underline the phenomenon that sooner or later will lead us to pass by the social contract era.

References.